\documentclass[twocolumn,showpacs]{revtex4}

\usepackage{graphicx}%
\usepackage{dcolumn}
\usepackage{amsmath}

\begin{document}

\title{Quantum cryptography using balanced homodyne detection}

\author{T. Hirano}
\email{hirano@qo.phys.gakushuin.ac.jp}
\author{T. Konishi}
\author{R. Namiki}
\affiliation{
Department of Physics, Gakushuin University, 
and CREST, Japan Science and Technology Corporation(JST),\\
Toshima-ku, Tokyo, 171-8588, Japan}

\date{\today}

\begin{abstract}
We report an experimental quantum key distribution that utilizes balanced homodyne detection, instead of photon counting, to detect weak pulses of coherent light. 
Although our scheme inherently has a finite error rate, it allows high-efficiency detection and quantum state measurement of the transmitted light using only conventional devices at room temperature.
When the average photon number was 0.1, an error rate of 0.08 and "effective" quantum efficiency of 0.76 were obtained.
\end{abstract}


\maketitle

According to quantum mechanics, one cannot obtain information about a single quantum system without disturbing its state \cite{bennett92} nor can one clone an unknown state \cite{wootters82}. 
Quantum cryptography is a technique for realizing secure communications exploiting these principles. 
The most popular protocol is quantum key distribution (QKD) in which two non-orthogonal states (B92 protocol) \cite{B92} or four states (BB84 protocol) \cite{BB84} are sent via a quantum channel in order to generate random keys owned only by the legitimate sender (usually called Alice) and the receiver (Bob).
These keys are then used to encode messages.

In practice, a faint laser pulse is usually used as the quantum system, and keys are encoded by its polarization or its phase.
Ideally, a single photon is desirable, but it is very difficult to generate it experimentally.
Most of the previous experimental and theoretical studies on QKD used or postulated photon counting as a means to detect weak light.
However, the usage of photon counting gives rise to two limitations. 
One is a technical limitation that at present there exists no efficient photon counter for infrared light, especially for 1.55-$\mu$m where optical loss in an optical fiber is minimum.
State-of-the-art experiments used a specially designed photon-counting system made up of cooled avalanche photo diode operated in a gated Geiger mode \cite{Marand1995,Hughes2000,Zbinden1998}.
For example, a quantum efficiency of 7\% for 1.55$\mu$m with a dark-count probability of 10$^{-4}$ per 2.6-nsec time-window was reported \cite{Ridordy1998}.
Another limitation is a more fundamental: the quantum state of the transmitted light cannot be directly measured; the state alternation is inferred only from the change of the error rate.
For example, when the eavesdropper (usually called Eve) changes the photon number distribution of the transmitted light while keeping the polarization (or the phase) and the mean photon number, Bob cannot notice the presence of Eve.
This feature allows Eve to perform many kinds of attacks and leads to security holes (one example is the photon number splitting attack \cite{Brassard}).

In this paper, we propose using balanced homodyne detection for implementing the BB84 protocol with phase coding \cite{pct}.
As we will explain, the above limitations associated with photon counting can be resolved by using balanced homodyne detection.
In order to demonstrate the experimental feasibility of our scheme, we have performed QKD by sending light pulses at 1.55-$\mu$m wavelength through an optical fiber of 20-cm length.
When the average photon number was 0.1, a bit-error-rate (BER) of 0.08 and "effective" quantum efficiency of 0.76 were obtained (effective detection efficiency was 0.076).

Balanced homodyne detection (sometimes called quadrature phase homodyne measurement) is a well-established quantitative method for measuring the quadrature-amplitude operator of the radiation field \cite{Yuen1983,Leonhardt}.
It has been developed as a means of detecting reduced quadrature-amplitude fluctuations (squeezed states of light) \cite{squeezing}.
In this method, a weak signal field interferes with a strong local oscillator (LO) on a beam splitter, and the difference of the intensities of the two outputs of the beam splitter is measured.
When the LO is much stronger than the signal, the output of the balanced homodyne detector is proportional to the quadrature-amplitude operator of the signal field \cite{Leonhardt}.
This is a very efficient method for measuring the quadrature amplitude of the signal, although no information on the conjugate quadrature is obtained.

In 1993, Smithey {\it et al.} demonstrated the determination of the Wigner distribution and the density matrix of a light field by measuring not only the variances but also the distributions of the quadrature amplitude of the field.
They used the so-called optical homodyne tomography (OHT) method in which the inverse Radon transform of the distributions was calculated.
A measurement of the density matrix provides all knowable information allowed by quantum mechanics.
For example, photon number statistics can be calculated from the density matrix.
Of course, it is impossible to measure the density matrix of a {\em single} quantum system; we need the ensembles of the same quantum state.
Such ensembles can be prepared by randomly switching the QKD and OHT procedure at the cost of a lowered transmission rate.
However, even without the OHT, distributions of the quadrature amplitude for some phases are obtained in the QKD procedure, thus limiting the range of allowable eavesdropping strategies.

Note that if a QKD protocol uses a phase coding relative to bright light as proposed in \cite{B92,Huttner1995}, Eve as well as Bob can perform balanced homodyne detection, although such possibilities have seldom been discussed so far.
Also from this viewpoint, it is important to investigate QKD using balanced homodyne detection.
Recently, several authors have presented quantum cryptography using homodyne detection \cite{Hillery,Ralph,Reid}.
These proposals use EPR-type correlation or squeezed states.
Our scheme uses only coherent states so that experimental realization is much easier.

We will now explain our implementation of QKD in more detail.
The protocol is basically an interferometric QKD using four nonorthogonal states \cite{B92}.
Laser pulses are split by an unsymmetrical beam splitter into two arms of a Mach-Zehnder interferometer.
Pulses in one arm, which we will call the local oscillator (LO), contain many photons (typically $n_{LO}\sim 10^6$ photon/pulse), and pulses in the other arm, which we will call the signal, contain photons at quantum level ($n_{sig}\leq 1$ photon/pulse).
Alice applies a random $\phi_A$=0$^{\circ}$-, 90$^{\circ}$-, 180$^{\circ}$-, and 270$^{\circ}$-phase shift to the signal; Bob, a random $\phi_B$=0$^{\circ}$-, 90$^{\circ}$-phase shift to the LO.
Bob, then, performs balance homodyne detection.
To put it concretely, Bob combines the signal field with the LO field by a 50-50 beam splitter.
Two photodiodes are used to monitor the intensities from two output ports.
Finally, the two photodiode outputs are subtracted, and the difference of photoelectrons $N_{\phi}$ is measured.
We denote the total phase shift between the signal and the LO by $\phi=\phi_A -\phi_B$.

\begin{figure}
\includegraphics[width=1\linewidth,clip=true]{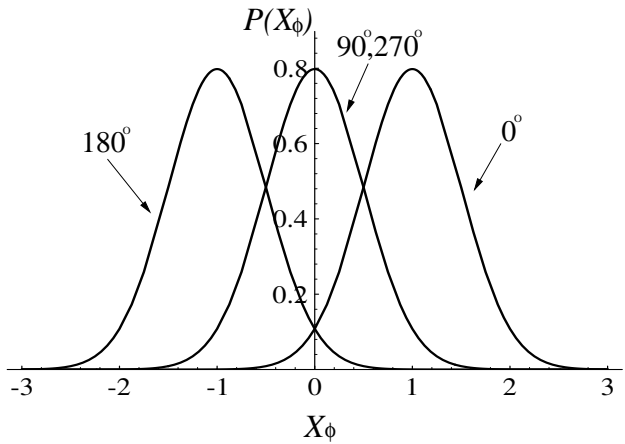}
\caption{
Theoretical probability distributions of the quadrature amplitude for total phase shifts are 0$^{\circ}$, 90$^{\circ}$, 180$^{\circ}$, and 270$^{\circ}$.
The signal photon number is 1 photon/pulse.}
\label{dist}
\end{figure}

The normalized quadrature amplitude of the signal 
$\hat{X} = (\hat{a}_{sig} + \hat{a}^{\dagger}_{sig})/2$ 
is obtained by $X_{\phi} = N_{\phi}/ 2\sqrt{n_{LO}}$, 
where $a_{sig}$ is the annihilation operator of the signal.
For each pulse, $X_{\phi}$ takes a random value due to quantum fluctuations.
Theoretically, the probability distribution $P(X_{\phi})$ is given by integrating the Wigner distribution over the conjugate variable $X_{\phi+90^{\circ}}$ \cite{Vogel}.
When the signal is in a coherent state, $P(X_{\phi})$ is given by a Gaussian function with a standard deviation of 1/2.
Figure \ref{dist} shows $P(X_{\phi})$ for $\phi$=0$^{\circ}$, 90$^{\circ}$, 180$^{\circ}$, and 270$^{\circ}$ when the average signal photon number is 1.
The probability distribution for $\phi$=90$^{\circ}$, $P(X_{90})$, and that for $\phi$=270$^{\circ}$ ,$P(X_{270})$, are the same, so it is impossible to differentiate them; in this case, Bob selected the wrong basis.
It is, however, possible to differentiate $\phi$=0$^{\circ}$ from $\phi$=180$^{\circ}$.
To do this, Bob sets up two threshold values $X_+$ and $X_-$ where $X_{-}\leq X_{+}$.
In the following, we set $X_{-}=-X_{+}$.
If the measured quadrature amplitude $X_{\phi}$ is larger than $X_+$ (in this case, we say that Bob's result is positive), Bob judges that $\phi$=0$^{\circ}$.
If $X_{\phi}$ is smaller than $X_-$ (Bob's result is negative), Bob judges that $\phi$=180$^{\circ}$.
Finally, if $X_{\phi}$ is between $X_-$ and $X_+$ (Bob gets an inconclusive result), Bob abandons the judgement.
Note that because $P(X_0)$ overlaps $P(X_{180})$, Bob's judgement is not always true.
This intrinsic bit error rate $e_{int}$ is the probability that $\phi$ is actually  180$^{\circ}$ even when Bob's result is positive, or $\phi$=0$^{\circ}$ for Bob's negative result.
The larger $X_+$ is the smaller $e_{int}$ becomes, but at the same time the probability, $p_{inc}$, that Bob gets inconclusive results becomes larger.
We define "effective" detection efficiency $p_d$ as the probability that Bob gets positive or negative results ($p_d=1-p_{inc}$).
A remarkable feature of our implementation of QKD is that both $e_{int}$ and $p_d$ are functions of the $n_{sig}$ and $X_{+}$.
The values of $e_{int}$ and $p_d$ can be easily calculated by using the error function.
For example, when $n_{sig}$=1 and $X_+$=$X_-$=0, $e_{int}$=0.023 and $p_d$=1.
If we choose $X_+$=$-X_-$=0.5, $e_{int}$ is greatly reduced to 0.0016 while $p_d$=0.84 changes a little.
When $n_{sig}$=0.1 and $X_+$=$-X_-$=1, $e_{int}$=0.047 and $p_d$=0.090.
In order to compare the performance of our scheme to that of the photon counting scheme, we define "effective" quantum efficiency $\eta_d$=$p_{d}/n_{sig}$.
In the last case, therefore, $\eta_d$=0.90.
These values demonstrate the excellent performance of balanced homodyne detection.
For a given value of $n_{sig}$ there exists an optimal $X_+$($X_-$) that maximizes 
the product of the effective detection efficiency and
the mutual information between Alice and Bob 
\cite{namiki}.

After an appropriate number of pulses have been transferred, Bob tells Alice which phase shift he applied for each pulse. Alice, then, tells Bob which phase shifts were correct. The correctly measured data is interpreted as a binary sequence according to the coding scheme ($\phi_A$=0$^{\circ}$ or 90$^{\circ}$)=1 and ($\phi_A$=180$^{\circ}$ or 270$^{\circ}$)=0 for Alice, and (positive result)=1 and (negative result)=0 for Bob. Finally, Alice and Bob perform error correction and privacy amplification procedures.

Let us now consider the security of this system.
Here we stress the advantage of the homodyne scheme that permits the measurement of quadrature amplitude distributions.
A detailed analysis is presented in a separate paper \cite{namiki}.
In the intercept-resend eavesdropping strategy, Eve intercepts selected light pulses and measures them, and then sends an appropriate state to Bob.
In the photon counting scheme, she can send a vacuum state (send nothing) for some pulses and send a higher intensity state for remaining pulses if the net rate of detection by Bob is unchanged \cite{BB84}.
In the homodyne scheme, however, this is not the case because such intervention changes the probability distribution of the measured quadrature amplitudes. 
In order to detect such changes, Bob graphs two probability distributions: one is for pulses for which he selected the correct basis ($\phi$=0$^{\circ}$ or 180$^{\circ}$), the other is for pulses for which he selected the wrong basis ($\phi=90^{\circ}$ or 270$^{\circ}$).
If, for example, Eve sends vacuum pulses, Bob's probability distribution for the correct basis becomes a superposition of Gaussians not only centered at $\pm \sqrt n_{sig}$ but also centered at 0 \cite{remark}.
The resolution of the quadrature-amplitude distribution becomes higher as the number of pulses increases.
Therefore, allowable eavesdropping strategies will be limited.
Note that since nonlinear optical processes, such as amplification or parametric processes, generally change the quantum state of light, sophisticated eavesdropping strategies would be detectable.

For Eve's intercept-resend eavesdropping to succeed, Eve must determine Alice's phase shift with high accuracy.
It is clear, however, that, if $n_{sig}$ is sufficiently small, Eve cannot do that.
She may take primarily two measurement schemes: she may measure the signal in the basis of her choosing as Bob does, or she may measure it in both bases by splitting it into two parts using a 50-50 beam splitter and measuring with 0$^{\circ}$-phase shift at one output and with 90$^{\circ}$-phase shift at the other.
For the former case, she obtains no information for half of pulses.
For the latter case, because the simultaneous measurement of two noncommuting observables introduces extra noise \cite{yamamoto-haus,Leonhardt}, her error rate is larger than the one for the single basis measurement. 
For example, when $n_{sig}$=1, the probability that she correctly differentiates between four phase shifts is only 0.708 (we assume that she must always choose one phase shift) \cite{namiki}.

If there is an optical loss in the communication channel (this is always true for long distance transmission), Eve can, in principle, replace the original channel with a more transparent one, and then put a beam splitter on the channel.
This beamsplitting attack does not change the state of the transmitted light so Bob cannot detect the presence of Eve.
Moreover, because Eve can stay closer to Alice than Bob, the signal intensity at Eve's port will be higher than the one at Bob's port.
However, even when Eve's signal-to-noise ratio is better than Bob's, Eve cannot get full information due to quantum fluctuations.
In addition, Bob can improve his error rate by discarding pulses (by increasing $X_+$). Eve, on the other hand, cannot improve her error rate for Bob's selected pulses because Eve's measurement result is uncorrelated with Bob's due to quantum fluctuations entering from the dark port of the beam splitter.
Therefore, Eve's Shannon information on the final key could be reduced to an arbitrarily small value by privacy amplification.

\begin{figure}
\includegraphics[width=1\linewidth,clip=true]{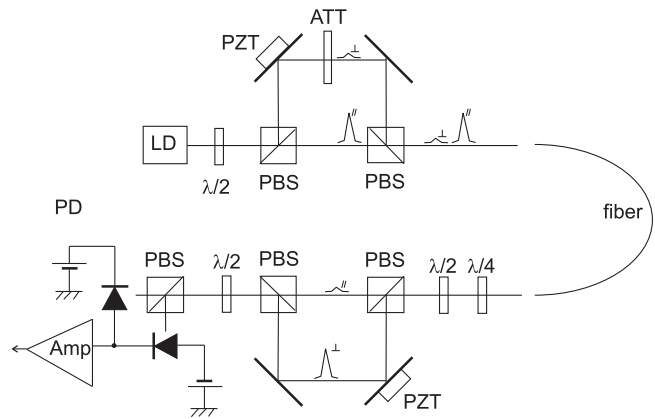}
\caption{
Experimental setup. We used a four-state protocol (BB84) in which keys are encoded in the phase difference between two pulses: (0$^{\circ}$,180$^{\circ}$) or (90$^{\circ}$, 270$^{\circ}$). In the transmission fiber, two pulses have mutually orthogonal polarization and are delayed by 500-psec. The light source is a 1.55-$m$m-wavelength semiconductor laser.}
\label{setup}
\end{figure}

In order to demonstrate the feasibility of our scheme, we performed a prototype QKD experiment.
The experimental setup is shown in Fig. \ref{setup}.
The light source is a 1.55-$\mu$m-wavelength semiconductor laser (Pico-Quanta model PDL-800) that generates $\sim$100-ps-duration pulses whose repetition rate can be controlled by a PC.
The laser output is split by a half-wave plate ($\lambda$/2) and a polarizing beam splitter (PBS) into the signal and LO pulse.
The signal pulse is attenuated to a single-photon level, and its phase can be varied by moving a mirror with a piezoelectric transducer (PZT).
We use time and polarization division to separate the signal and the LO in a transmission fiber \cite{Marand1995}.
A quarter-wave plate ($\lambda$/4) and a half-wave plate are used to control the polarization of the signal and the LO after they traverse the fiber.
The signal and LO overlap each other in time on the second PBS.
Their polarizations are then rotated by 45$^{\circ}$ so they interfere on the third PBS.
The balanced detector consists of two InGaAs PDs (Hamamatsu photonics model G3476-03) and a charge-sensitive amplifier (Digitex model HIC-1576 \cite{taniguchi}).
The quantum efficiency of the PD, $\eta_{PD}$, was measured to be 0.85$\pm$0.03.
The detector output is recorded by an A/D board installed in a PC.
The overall electronic detection noise was 1010 electrons rms.
The phase drift of our interferometer was less than $\lambda$/10 in 10-seconds.
The fringe visibility $V$ was 0.8 and was limited by imperfect spatial overlap between the signal and the LO.

\begin{figure}
\includegraphics[width=1\linewidth,clip=true]{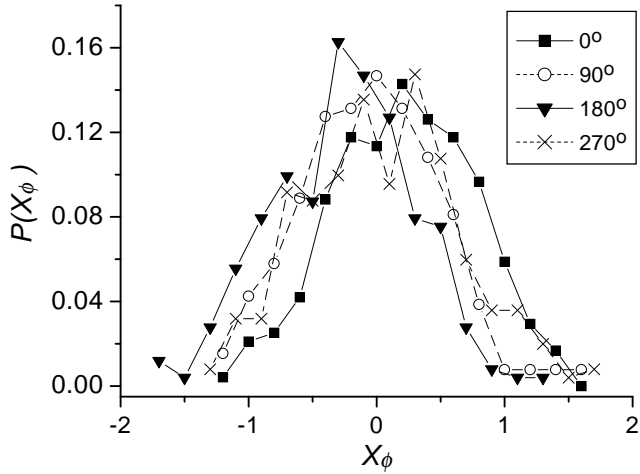}
\caption{
The measured distributions of the quadrature amplitude of the signal for different total phase-shifts between the signal and the LO pulse (0$^{\circ}$, 90$^{\circ}$, 180$^{\circ}$, and 270$^{\circ}$). Total number of pulses is 10$^3$.
The signal intensity was 0.1 photons/pulse, the LO intensity was 2x10$^6$ photons/pulse, and the fringe visibility of the interferometer was 0.8 (limited by spatial overlap).}
\label{distributions}
\end{figure}

Figure \ref{distributions} shows the measured $P(X_{\phi})$ for $\phi$=0$^{\circ}$, 90$^{\circ}$, 180$^{\circ}$, and 270$^{\circ}$.
$n_{sig}$ was 0.1 photons/pulse, and $n_{LO}$ was 2x10$^6$ photons/pulse.
There was a total of 10$^3$ pulses.
It took about 1-min to send 10$^3$-pulses; the repetition rate of the laser was set very low because the response time for moving a mirror with PZT was slow.
Use of electro-optical modulator will enable us to increase the repetition rate and decrease the phase error.
The mean and the standard deviations of the distributions are shown in Table 1.
The measured values of the mean are consistent with theoretical values of $\pm\sqrt{n_{sig}\eta_{PD}}V$=$\pm$ 0.23.
The standard deviations are large than the ideal value of 0.5 mainly due to the amplifier noise.

\begin{table}
\caption{Mean and standard deviations of measured $P(X_{\phi})$}
\begin{tabular}{ccc}
\hline\hline
\makebox[0.2\linewidth]{$\phi$} & 
\makebox[0.25\linewidth]{Mean} & 
\makebox[0.43\linewidth]{Standard deviation} \\ \colrule
0$^{\circ}$ & 0.230$\pm$ 0.022 & 0.585$\pm$ 0.023 \\
90$^{\circ}$ & -0.060$\pm$ 0.018 & 0.552$\pm$ 0.018 \\
180$^{\circ}$ & -0.227$\pm$ 0.038 & 0.562$\pm$ 0.038 \\
270$^{\circ}$ & 0.022$\pm$ 0.050 & 0.0634$\pm$ 0.051 \\
\hline\hline
\end{tabular}
\end{table}

QKD was performed using the same data as shown in Fig. \ref{distributions}.
Bob selected the correct basis ($\phi=0^{\circ}$ or 180$^{\circ}$) for 490 pulses.
For $X_{+}$=0, $p_d$=1 and BER was 0.34.
For $X_{+}$=0.98, $p_d$=0.076 ($\eta_d$=0.76) and BER was 0.081.
Although these values are worse than the ideal estimation shown above, 
they still demonstrate the superior performance of our scheme, 
and will be improved in the future by using a fiber-optics-based 
interferometer, a low-noise, high-speed, charge-sensitive amplifier, and so on.
Note that because for Eve $n_{sig}$=1 is already too low to effectively perform eavesdropping and for Bob $n_{sig}$=0.1 is high enough to perform QKD, our scheme would be feasible in the presence of 10-dB optical loss.

In conclusion, we have presented a novel scheme for QKD that utilizes balanced homodyne detection in the phase-encoding BB84 protocol.
The scheme has an inherently finite error rate, but previous schemes also have finite error rates in practice.
The advantage of our scheme is that measuring the distributions of quadrature amplitudes limits the allowable eavesdropping strategies.
In addition, if Alice randomly adds auxiliary phase-shift, the density matrix of the signal could be measured by optical homodyne tomography.
We have performed a prototype QKD experiment at 1.55-$\mu$m wavelength.
When the average photon number was 0.1, a BER of 0.08 and "effective" quantum efficiency of 0.76 were obtained (effective detection efficiency was 0.076).

We thank M. Koashi, A. Shimizu, T. Kuga, and Y. Torii for helpful discussions, and K. Komori and M. Yanagihara for work during the early stage of the experiment.
T. H. acknowledges the financial support from the Research Foundation for Opto-Science and Technology.

\end{document}